\documentclass[runningheads]{llncs}
\usepackage[T1]{fontenc}
\def\doi#1{\href{https://doi.org/\detokenize{#1}}{\url{https://doi.org/\detokenize{#1}}}}

\usepackage{amsmath,amssymb}
\usepackage{graphicx}
\usepackage{listings}
\usepackage{todonotes}

\usepackage{amsmath,amsfonts,bm}

\def\eqref#1{equation~\ref{#1}}

\def\1{\bm{1}}

\def\vmu{{\bm{\mu}}}

\def\va{{\bm{a}}}

\def\vc{{\bm{c}}}

\def\vx{{\bm{x}}}
\def\vy{{\bm{y}}}

\def\mA{{\bm{A}}}

\def\mI{{\bm{I}}}

\def\mR{{\bm{R}}}

\def\mW{{\bm{W}}}

\def\mLambda{{\bm{\Lambda}}}

\DeclareMathAlphabet{\mathsfit}{\encodingdefault}{\sfdefault}{m}{sl}
\SetMathAlphabet{\mathsfit}{bold}{\encodingdefault}{\sfdefault}{bx}{n}

\begin{document}
\setlength{\abovedisplayskip}{10pt}
\setlength{\belowdisplayskip}{10pt}
\setlength{\abovedisplayshortskip}{10pt}
\setlength{\belowdisplayshortskip}{10pt}
\title{NUQ: A Noise Metric for Diffusion MRI via Uncertainty Discrepancy Quantification} 
\titlerunning{NUQ: A Noise Metric for Diffusion MRI}
%
\author{Shreyas Fadnavis\inst{1} \and
Jens Sj\"olund\inst{2} \and
Anders Eklund\inst{3, 4, 5} \and 
Eleftherios Garyfallidis\inst{1}}
\authorrunning{Fadnavis et al.}
%
\vspace{-2mm}
\institute{Intelligent Systems Engineering, Indiana University Bloomington, USA \and
Department of Information Technology, Uppsala University, Sweden \and 
Department of Biomedical Engineering, Link\"oping University, Sweden \and
Center for Medical Image Science \& Visualization (CMIV), Link\"oping University, Sweden \and
Division of Statistics \& Machine Learning, Department of Computer \& Information Science, Link\"oping University, Sweden}

\maketitle              
\vspace{-7mm}
\begin{abstract}
Diffusion MRI (dMRI) is the only non-invasive technique sensitive to tissue micro-architecture, which can, in turn, be used to reconstruct tissue microstructure and white matter pathways. The accuracy of such tasks is hampered by the low signal-to-noise ratio in dMRI. Today, the noise is characterized mainly by visual inspection of residual maps and estimated standard deviation. However, it is hard to estimate the impact of noise on downstream tasks based only on such qualitative assessments. To address this issue, we introduce a novel metric, Noise Uncertainty Quantification (NUQ), for quantitative image quality analysis in the absence of a ground truth reference image. NUQ uses a recent Bayesian formulation of dMRI models to estimate the uncertainty of microstructural measures. Specifically, NUQ uses the maximum mean discrepancy metric to compute a pooled quality score by comparing samples drawn from the posterior distribution of the microstructure measures. We show that NUQ allows a fine-grained analysis of noise, capturing details that are visually imperceptible. We perform qualitative and quantitative comparisons on real datasets, showing that NUQ generates consistent scores across different denoisers and acquisitions. Lastly, by using NUQ on a cohort of schizophrenics and controls, we quantify the substantial impact of denoising on group differences.
\vspace{-2mm}
\end{abstract}
\vspace{-5mm}
\section{Introduction}
\begin{figure}[ht]
\centering
\includegraphics[width=1\textwidth]{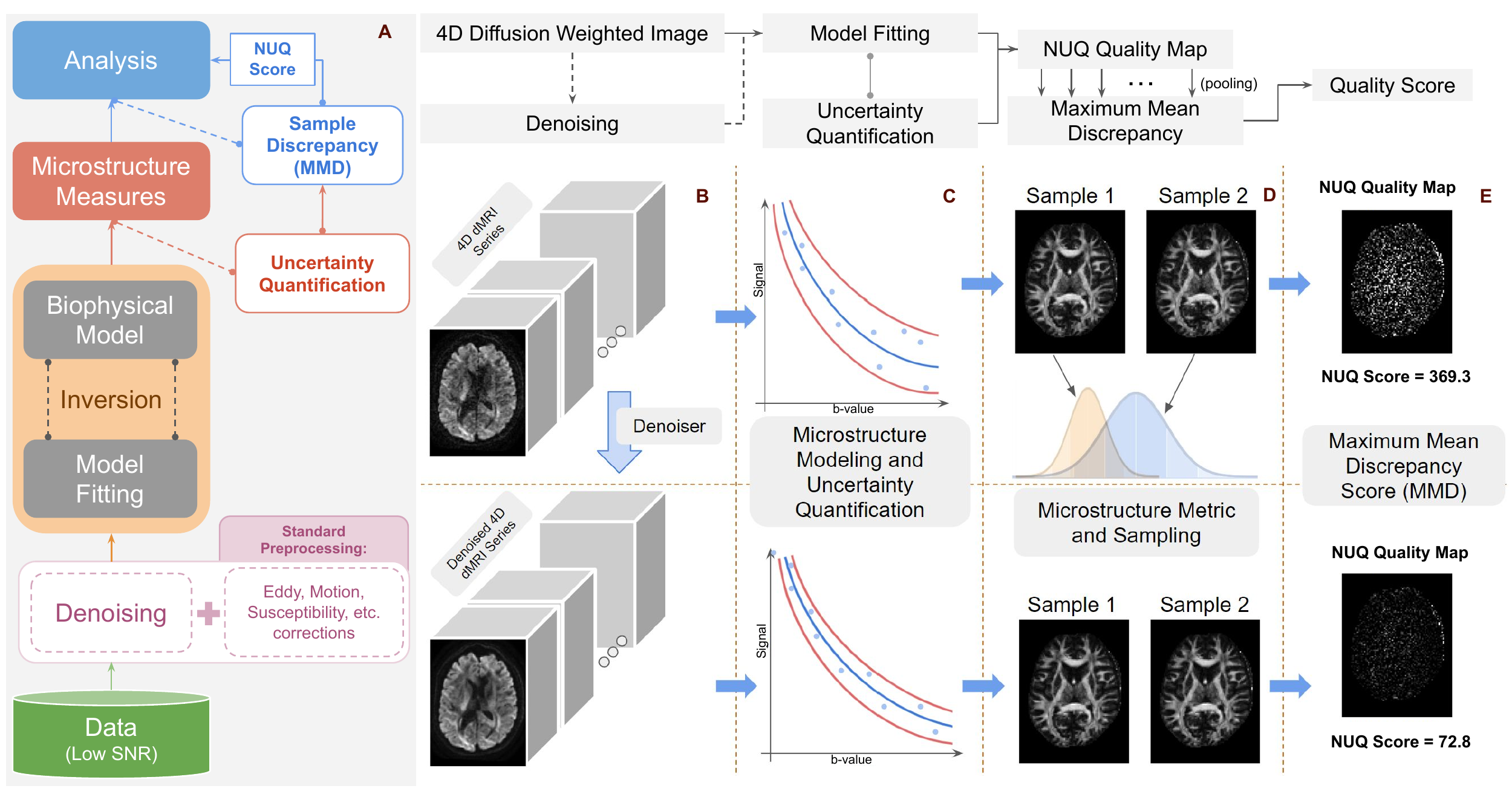}
\caption{A) Typical dMRI analysis workflow depicting points where NUQ is plugged-in. B) A 4D series of diffusion-weighted images before and after denoising. C) Comparison showing that denoising reduces uncertainty in the model fit. D) Two independent draws of FA from the posterior distributions of both noisy and denoised data. E) NUQ quality maps and scores computed before and after denoising.}
\label{fig:nuq_flow}
\end{figure}

\vspace{-2mm}
Diffusion MRI (dMRI) has proven to be an invaluable tool for noninvasive mapping of the tissue microstructure and reconstruction of white matter fascicles. However, dMRI data is prone to low signal-to-noise ratio acquisitions owing to a pursuit of higher resolution and practical restrictions on acquisition times \cite{hutchinson2017analysis}. Most of the downstream clinical analysis in dMRI relies on signal reconstruction and modeling \cite{novikov2018modeling}. A typical workflow for dMRI analysis, as shown in Fig.~\ref{fig:nuq_flow}A, starts with denoising and other artefact corrections (such as Eddy, Head Motion, Susceptibility and others \cite{tax2022s}) which dMRI data are commonly prone to. The next step requires solving an inverse problem to characterize tissue microstructure \cite{novikov_fieremans_jespersen_kiselev_2018}. Typically this involves fitting either a phenomenological \cite{basser_mattiello_lebihan_1994,jensen_helpern_2010,pasternak2009free,zhang2012noddi} or a mechanistic model \cite{aganj_lenglet_sapiro_yacoub_ugurbil_harel_2009,ozarslan2013mean,tournier_calamante_connelly_2007} to the data. The model is then used to infer the underlying (sub-voxel resolution) tissue microstructure. Due to the inherent noise in the acquired signal, this ill-posed inverse problem leads to fitting degeneracies and unreliable estimates of the underlying cellular organization \cite{fadnavis_p2s}. These errors propagate and amplify the uncertainty in downstream tasks such as tractography and group-level analyses. In this work, we strictly focus on the effects that noise and subsequent denoisers have on downstream tasks to avoid confounding smoothing effects from other preprocessing steps.

Denoising plays a crucial role in the dMRI analysis pipeline, and has been tackled from a variety of perspectives over the years. Most approaches fit into one of two main categories: denoisers that work with magnitude data only, and those that use the complex signal (to convert the Rician noise behaviour to Gaussian and address the noise floor issue) \cite{koay_ozarslan_basser_2009}. We consider the first case of denoisers that deal with magnitude data. These can be further categorized based on their underlying assumptions; popular and successful ones include low-rank approximations (Marchenko Pastur PCA) \cite{veraart_novikov_christiaens_ades-aron_sijbers_fieremans_2016}, self-similarity (Multi-resolution Optimized Non-Local Means (Multi-ONLM)) \cite{coupe_manjon_robles_collins_2012} and statistical independence (Patch2Self) \cite{fadnavis_p2s}. To demonstrate the utility of the proposed noise metric---Noise Uncertainty Quantification (NUQ)---we show that it captures subtle details across this range of denoisers as well as with data acquired with different schemes. In particular, we perform experiments with a diffusion tensor imaging (DTI) model \cite{basser_mattiello_lebihan_1994} on single-shell data and a mean apparent propagator (MAP) MRI model \cite{ozarslan2013mean} on multi-shell data. 
\vspace{-4mm}
\subsection{Related Work}
\label{sec:rel_work}
\vspace{-1.5mm}
\textbf{Image Quality Analysis:} Quantification of noise and its effect on the signal remains an open problem in dMRI data. Most  methods for analyzing noise rely on visual inspection of residuals (computed as the mean squared distance between the raw and denoised image) per 3D volume/ gradient direction or by computing a voxel-wise standard deviation map \cite{veraart_novikov_christiaens_ades-aron_sijbers_fieremans_2016}. Neither of these approaches allow an objective quality assessment, nor do they reflect the impact on model fitting or how to incorporate it in group-level analyses. Several different image quality metrics have been proposed in the image processing literature, including structural similarity \cite{wang2004image}, natural scene statistics \cite{wang2011reduced}, and visual attention \cite{engelke2011visual}. Arguably, the most widely used out of these is the structural similarity index metric (SSIM) \cite{wang2004image}. SSIM mainly aims to separate structural components of the signal from the non-structural fluctuations and distortions. By doing so, it penalizes different parts of the image with different regularization strengths. SSIM then generates a quality map from which information is pooled to compute an image-level quality score as a single scalar. 

We propose a new metric, NUQ, which is tailored to capture the requirements of medical image quality analysis. Unlike SSIM and other approaches mentioned above, NUQ uses uncertainty in the model's fit to the dMRI data to capture the effect of noise so that it can be used in the downstream subject and group level analyses. NUQ takes advantage of the fact that most modeling approaches used in dMRI use linear combinations of nonlinear/orthogonal-basis functions and are solved by least-squares optimization routines. Instead of getting a single point estimate of the model parameters fit to the data, we use a Bayesian probabilistic model to derive a posterior distribution of the model parameters \cite{Sjolund2018}. While we make use of a Bayesian framework (explained in detail in Sec.~\ref{sec:buq}), NUQ can be used in combination with any generative model as it only requires samples from the posterior distribution. Samples of the microstructure measures (derived via fitted parameters) are drawn from this posterior. These samples are then compared using a two-sample testing procedure to compute the discrepancy among them. The \textit{key idea} of the NUQ setup is that higher noise levels lead to higher uncertainty (variance) in the parameter estimates. 
Thus samples from a noisier distribution have a higher discrepancy than its denoised counterpart.
\\
\textbf{Maximum Mean Discrepancy:} As mentioned above, a key component of our noise metric is to compute the discrepancy between samples from the posterior distribution. 
Many metrics for two-sample testing have been proposed; famous ones include the Kolmogorov-Smirnov test \cite{berger2014kolmogorov}, Wasserstein distance \cite{ramdas2017wasserstein} and the Maximum Mean Discrepancy Metric (MMD) \cite{gretton12a}. One could also use the Kullback-Leibler Divergence but this is not a proper metric (since it is asymmetric) and, besides, requires binning of the data. Instead, to compute the NUQ metric, we favor the Maximum Mean Discrepancy (MMD) which allows for kernelized two-sample testing. In addition to projecting the data into a higher dimensional feature space, MMD provides capabilities to compute goodness-of-fit (of model to data) and measure statistical independence (between samples). The MMD is an integral probability measure which maximizes the difference between the mean function values of the two distributions $p(\vx)$ and $q(\vx)$ under comparison. MMD \textit{transforms a sample from each distribution into higher order features} using a kernel from a reproducing kernel Hilbert space $\mathcal{H}$ that enforces smoothness. Therefore for $\mathbf{x} \sim P$, $\mathbf{y} \sim Q$ and given a kernel $\varphi$:
\begin{align*}
    \operatorname{MMD}^{2}(P, Q)
    =& \mathbb{E}_{\mathbf{x}, \mathbf{x}^{\prime} \sim \mathbf{P}} \left[\varphi\left(\mathbf{x}, \mathbf{x}^{\prime}\right)\right]+\mathbb{E}_{\mathbf{y}, \mathbf{y}^{\prime} \sim Q}\left[\varphi\left(\mathbf{y}, \mathbf{y}^{\prime}\right)\right]- \\
    & 2\mathbb{E}_{\mathbf{x} \sim P, \mathbf{y} \sim Q} \left[\varphi(\mathbf{x}, \mathbf{y})\right]    
\end{align*}
Here, the first two terms measure the similarity within the respective distributions, whereas the last term measures the similarity between the two distributions. 
Thus, the MMD score will be a finite, nonnegative, scalar quantity that equals  $0$ if and only if the distributions $P$ and $Q$ are equal. We propose to use MMD to summarize the uncertainty in estimates of microstructural metrics. It is particularly appealing for NUQ as it is a distribution-free approach, i.e. it does not make any assumptions about the underlying sample distributions.
\vspace{-4mm}
\section{Methods}
\vspace{-3.5mm}
\label{sec:buq}
The proposed NUQ metric is computed for a 4D series of diffusion-weighted image in three independent stages. The first stage consists of fitting a phenomenological (cumulative expansions of the signal) or mechanistic model (linear combinations of orthogonal basis functions of the signal) to the data. Most microstructure models for dMRI are fitted using a variation of the weighted least squares \cite{Sjolund2018}. NUQ can be extended to any of these models irrespective of the requirements different models may have on the type of data acquisition such as single-shell and multi-shell data. In the second stage, we estimate the uncertainty of the model fit to the data using a Bayesian framework. Lastly, in the third stage we pool the uncertainty estimates by computing discrepancy between the samples obtained from the posterior in the second stage.

\textbf{Uncertainty Quantification:} Many common dMRI models fit the measurements $\vy$, performed with experimental settings $\vx$, to a model defined as a linear combination of $d$ basis functions $\va(\vx)$ with coefficients $\vc$, i.e. $\vy \approx \sum_{j=1}^d c_j \va_j(\vx)$. Most of the dMRI literature has focused on fitting a single point estimate $\hat{\vc}$ of the coefficients using an instance of the least-squares expression:
\begin{equation}
    \hat{\vc} = \left(\mA^T\mW\mA + \mLambda\right)^{-1}\mA^T\mW\vy,
\end{equation}
where $\mA_{:,j}=\va_j(\vx)$, $\mW$ is a weight matrix and $\mLambda$ is a Tikhonov regularization matrix. Recently, it was shown that those estimates can be framed as merely the mean of a probability distribution over the coefficients \cite{Sjolund2018}. This class of linear models fitted with least-squares admit a particularly neat interpretation based on Bayesian statistics, resulting in a closed-form expression for the posterior:
\begin{equation}
p(\vc \mid \vy, \vx) = t_\nu(\vc ; \vmu, \mR),    
\end{equation}
where the mean $\vmu$, degrees of freedom $\nu$, and correlation matrix $\mR$ are given by
\begin{align}
    \vmu &= \left(\mA^T\mW\mA + \mLambda\right)^{-1}\mA^T\mW\vy,\\
    \nu &= \text{Tr}\left((\mI - \mA\vmu)\mW^{-1}(\mI - \mA\vmu)^T\right),\\
    \mR &= \frac{\nu-2}{\nu}\hat{\sigma}^2\left(\mA^T\mW\mA + \mLambda\right)^{-1},\\
    \hat{\sigma}^2 &= \frac{\|\vy - \mA\vmu\|_2^2}{\nu}.
\end{align}
The quantity $\hat{\sigma}^2$ is the estimate of the residual variance.
Typically, to analyze dMRI data via microstructure models, we want to estimate a scalar property $z$, defined as a possibly non-linear function of the coefficients, $z = f(\vc)$. Given that the coefficients follow a (posterior) probability distribution, this induces a probability distribution over the property $z$. We might not have access to this distribution in closed-form, but this is not an issue since NUQ only requires \emph{samples} from the posterior distribution. If it is easy to draw samples from the posterior over the coefficients, as in the linear models described above, then these can be passed through $f(\vc)$ to obtain samples from the posterior $p(z|y)$. In this work, we focus on commonly used properties $z$ from two widely used models, DTI \cite{basser_mattiello_lebihan_1994} and MAP-MRI \cite{ozarslan2013mean}, which from an application standpoint cover a wide spectrum of diffusion acquisition schemes (single and multi-shell) \cite{descoteaux_2015}. Note that NUQ can easily be extended to other advanced models such as QTI \cite{westin2016q} which require multidimensional diffusion encoding acquisitions \cite{Sjolund2015}. For the DTI model, we make use of the fractional anisotropy (FA) property which is the most common measure studied for dMRI data. For the MAP-MRI model, we use the return-to-origin probability (RTOP) measure \cite{ozarslan2013mean} which can be used for multi-shell data. While RTOP can be analytically computed as a function of the MAP-MRI model parameters, $f(\vc)$, FA is a non-linear combination of the DTI model parameters and therefore needs to be sampled from the posterior. 

\begin{figure}[ht]
\centering
\includegraphics[width=1\textwidth]{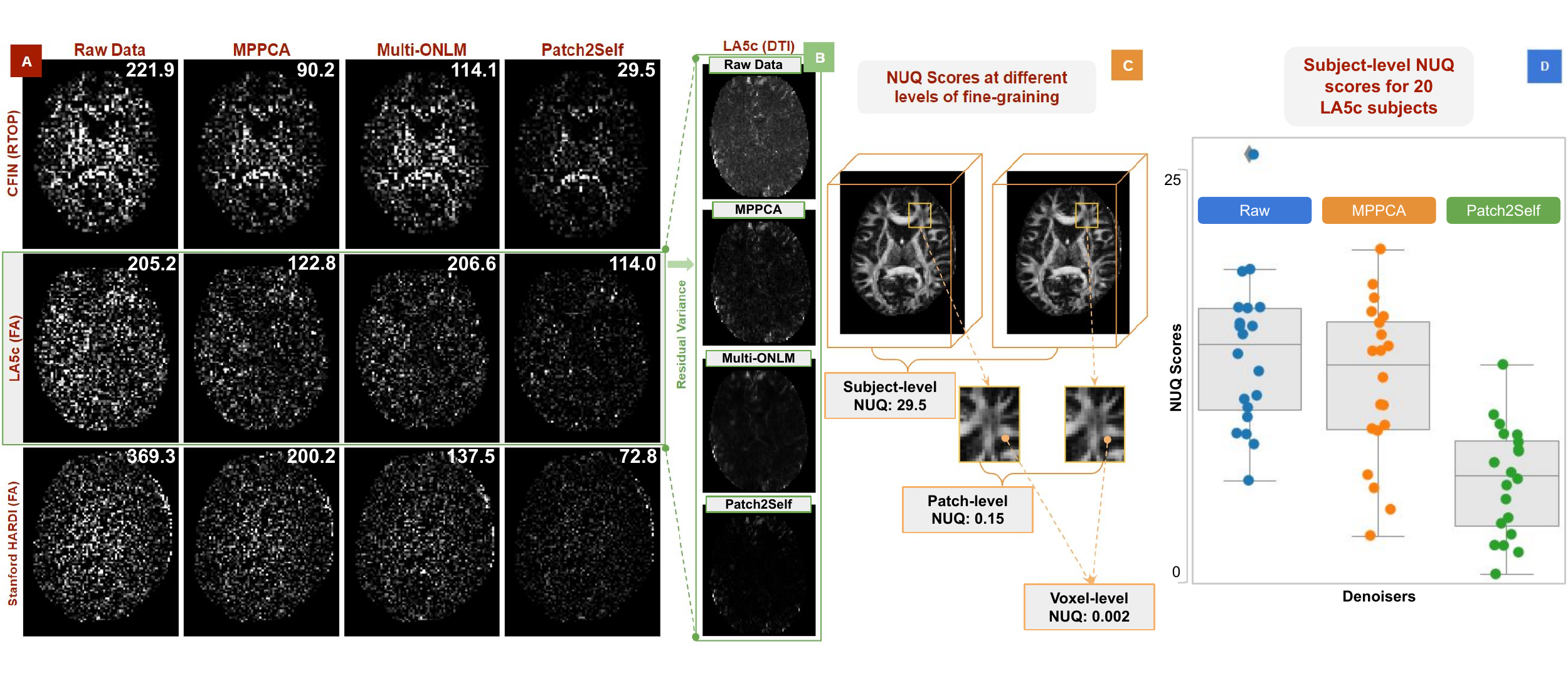}
\caption{A) Each row represents three different datasets and each column corresponds to a different denoiser outcome. The voxel-level NUQ is depicted via quality maps and the subject-level summary score is given at the top of it's corresponding map. B) Residual variance ($\hat{\sigma}^2$) estimated after denoising via different denoisers for the same LA5c subject as in (A). C) NUQ scores computed at different levels of fine-graining (subject, patch and voxel-level). D) Comparison of 20 subjects from the LA5c cohort via subject-level NUQ scores for raw data and data denoised using MPPCA and Patch2Self.}
\label{fig:nuq_multi}
\end{figure}

\textbf{Pooling Quality Scores using MMD:} While the the voxel-level uncertainty within NUQ is easy to compute by simply taking a difference between two samples drawn from the posterior, we also enable computing patch-level and subject level measures using the MMD metric (described in Sec.~\ref{sec:rel_work}). This allows fine-graining into the details in 2D/3D. For example, as shown in Fig.~\ref{fig:nuq_multi}C, one can compute the NUQ scores between two samples drawn for the posterior of the FA parameters for the entire subject or pick out particular slices/patches from the same subject and analyze those. To do so, we allow switching between different kernels $\varphi$ (such as radial basis function, polynomial kernel, Gaussian kernel) within the MMD metric used in NUQ. For 1D/3D data, we make use of the linear kernel and for 2D data we recommend using the polynomial kernel of degree $\geq 2$. One can also use the witness function \cite{gretton12a} to choose the kernel within MMD. In the supplement, we show that changing the degree of the polynomial does not affect consistency of NUQ scores. While we integrate MMD within the NUQ metric, we also provide the flexibility to switch to other metrics such as Wasserstein distance (see supplement for more details). Lastly, to enable deeper insight into the noise variance and effect of denoising performance, NUQ allows for estimation of the residual variance from the uncertainty of the model fit. As delineated in Eq.~6, the residual variance, $\hat{\sigma}^2$ $\propto \|\vy - \mA\vmu\|_2^2$, implying that as the model estimate approaches the posterior mean, $\mu$, the variance of distribution from which property $z$ is computed reduces. This directly reduces the estimated residual variance. See Fig.~\ref{fig:nuq_multi}B for instance, where the residual variance of an LA5c subject (sub-50020) \cite{poldrack2016phenome} reduces via denoising from different denoisers.
\\
\textbf{Properties of NUQ:} By design, NUQ can be applied to dMRI data from any acquisition scheme and anatomical location as long as it can be described by a microstructure model such as DTI. NUQ is also agnostic to the model used to fit the data and the optimization procedure used to do so as long as a generative model can be used to sample from the posterior. Owing to the MMD metric, the pooled NUQ scores (patch or subject-level) will always be nonnegative and finite. NUQ is strictly based on assessment of the model fit via uncertainty and is therefore not affected by smoothness. In our results, we show that the modeling uncertainty is unaffected even when using denoising methods known to smooth the data. 
\begin{figure}[ht]
\centering
\includegraphics[width=1\textwidth]{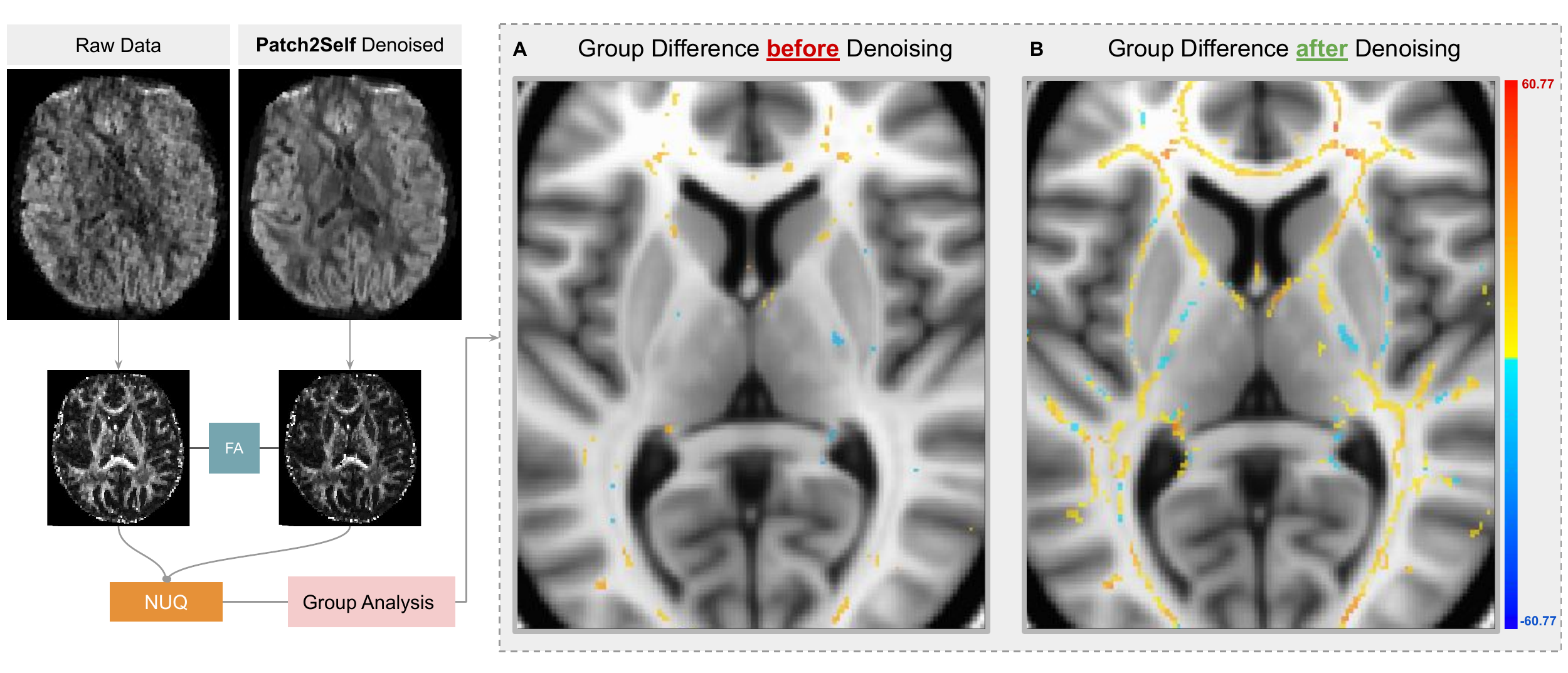}
\caption{Left panel shows the raw data and its Patch2Self denoised output. NUQ scores based on FA are computed from both which is then used to do group-analysis via TBSS. Right panel depicts the group difference between 20 schizophrenics and 20 controls before denoising in (A) and after denoising with Patch2Self in (B).}
\label{fig:nuq_tbss}
\end{figure}
\vspace{-4mm}
\section{Results}
\vspace{-4mm}
We first consider the effect of three different denoisers (MPPCA, Multi-ONLM and Patch2Self) on three datasets using different acquisition schemes. For each combination, we compute NUQ scores at a voxel-level (via NUQ quality maps) and subject-level as shown in Fig.~\ref{fig:nuq_multi}A. The CFIN data \cite{hansen2016data} has a multi-shell sampling scheme with 496 gradient directions and b-values ranging from 200-3000 $\text{s/mm}^2$. This data was fitted with the MAP-MRI model and the NUQ was estimated via the RTOP measure. Patch2Self yields the best performance with decreased uncertainty in the  NUQ quality maps and least subject level (NUQ: 29.5) score. Multi-ONLM (NUQ: 114.1) performs worse than MPPCA (NUQ: 90.2) resulting in the highest NUQ scores. A similar effect was seen on a subject from the LA5c study (sub-10506) acquired with a standard DTI protocol with 64 gradient directions at b-value 1000 $\text{s/mm}^2$ (second row)---here Patch2Self performed slightly better than MPPCA with Multi-ONLM yielding a score (NUQ: 206.6) worse than the raw data (NUQ: 205.2). This indicates that Multi-ONLM is not well suited for such data. In contrast, on the last dataset, Stanford HARDI data \cite{rokem_2016} with 150 gradient directions at b-value 2000 $\text{s/mm}^2$ and high-angular resolution protocol), Multi-ONLM (NUQ: 137.5) performed better than MPPCA (NUQ: 200.2), but still worse than Patch2Self (NUQ: 72.8). This comparison shows that NUQ generates consistent scores and results across acquisitions. To show that NUQ scores are consistent across subjects within the same dataset, we compared denoising performance on 20 randomly chosen subjects from the LA5c controls cohort, see Fig.~\ref{fig:nuq_multi}D. We see that Patch2Self achieves substantially lower NUQ scores than MPPCA across all subjects. Interestingly, we observe a higher NUQ score after denoising with MPPCA on a few subjects in comparison with raw data. NUQ is also \textit{not sensitive to spatial smoothing} as it only works with the uncertainty of derivative measures from the microstructure models. This can be seen in Fig.~\ref{fig:nuq_multi}A where different denoisers are compared via RTOP and FA. Multi-ONLM is known to cause spatial smoothing whereas MPPCA and Patch2Self do not \cite{fadnavis_p2s} (see supplement for a comparison). One can see qualitatively (via NUQ quality maps) and quantitatively (via NUQ scores) that for the case of RTOP on the CFIN data and FA on the LA5c data, despite smoothing the signal, Multi-ONLM has a higher uncertainty as compared to MPPCA and Patch2Self. Implicitly, NUQ allows assessing the suitability of a denoiser for a q-space sampling scheme or acquisition strategy.
\begin{figure}[ht]
\centering
\includegraphics[width=1\textwidth]{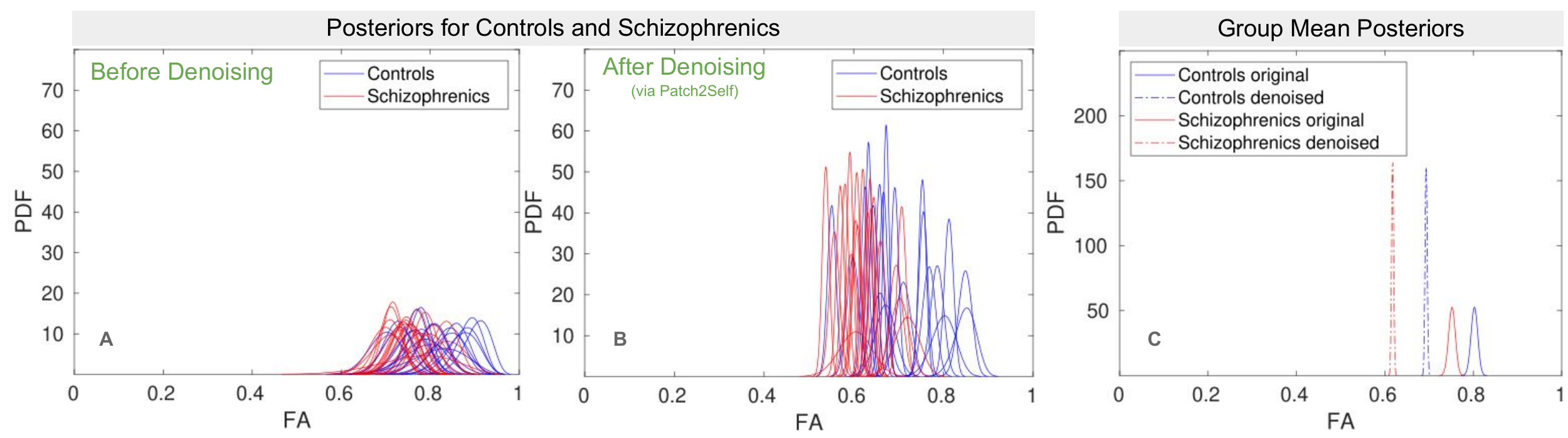}
\caption{Posteriors of randomly chosen voxels for 20 controls and 20 schizophrenics before (A) and after denoising (B) using NUQ scores computed on FA samples. C) Summarizes the results via posterior group means before and after denoising. }
\label{fig:nuq_beta_scz}
\end{figure}
\\
\textbf{Group-level Analyses:} To investigate how denoising affects a standard TBSS \cite{smith2006tract} group analysis, we used data from 20 controls and 20 schizophrenics from the LA5c dataset \cite{poldrack2016phenome}. Patch2Self denoising was applied to each diffusion dataset (since prior results indicated better performance), and the Bayesian modelling approach \cite{Sjolund2018} was then used to obtain 100 FA maps per subject (with and without denoising) as samples from the posterior. One FA map per subject was used for the standard steps in TBSS, to register and skeletonize all subjects to a common white matter skeleton. As the denoising can affect the registrations and the group skeleton, the obtained registrations and group skeleton for the denoised data were also used for the original data to transform the 100 FA maps per subject to the group skeleton. In each voxel in the skeleton mask, Bayesian group analyses (via Bayesian t-scores) were finally performed using original and denoised data to obtain posteriors of group mean and group difference. Each subject in this group analysis was weighted by the reciprocal of the standard deviation of its posterior, to downweight subjects with a higher NUQ scores. The obtained group difference posteriors in each voxel were converted to Bayesian t-scores by dividing the posterior mean by the posterior standard deviation. This is quantified at a group-level on the TBSS skeleton as shown in in Fig.~\ref{fig:nuq_tbss}. The larger the t-score, the more likely it is that controls have greater FA on average. A negative t-score means that schizophrenics have greater FA on average. As shown in Fig.~\ref{fig:nuq_tbss}, the group difference improves substantially after denoising based on FA values. Our results also abide by the findings of \cite{kubicki2007review} where the schizophrenics have reduced FA as compared to controls. The overall reduction in the FA after denoising is expected on account of the reduced Rician bias \cite{tax2022s} and reduced uncertainty explained in \cite{jones2004effect}. Comparing the group mean posteriors for raw and denoised data in randomly selected voxels shows a rather big difference, which can also be seen when looking at the posterior of each subject in the same voxels. The denoising substantially reduces the NUQ scores of the FA estimates, resulting in higher Bayesian t-scores. 
\vspace{-5mm}
\section{Conclusions}
\vspace{-3mm}
We developed a new metric, NUQ, that enables automated image quality assessment for dMRI data. Using a generative model via Bayesian uncertainty quantification, NUQ assesses the model fit of any microstructure model. Apart from deriving quality maps sensitive to noise, NUQ also scores the image at subject-, patch- or voxel-level. We showed how NUQ can be used to evaluate different denoisers within a subject and across a cohort. Furthermore, with the help of a cohort of 20 controls and 20 schizophrenics, we delineated how NUQ can be used to analyze group differences. A well-tested implementation of NUQ will be released as a part of DIPY software package (module: \texttt{dipy.stats.nuq}) to disseminate it to the community.
\bibliographystyle{plain}
\bibliography{nuq_citations.bib}

\begin{thebibliography}{10}

\bibitem{aganj_lenglet_sapiro_yacoub_ugurbil_harel_2009}
Iman Aganj, Christophe Lenglet, Guillermo Sapiro, Essa Yacoub, Kamil Ugurbil,
  and Noam Harel.
\newblock Reconstruction of the orientation distribution function in single and
  multiple shell q-ball imaging within constant solid angle.
\newblock {\em Magnetic Resonance in Medicine}, 2009.

\bibitem{basser_mattiello_lebihan_1994}
P.j. Basser, J.~Mattiello, and D.~Lebihan.
\newblock {MR} diffusion tensor spectroscopy and imaging.
\newblock {\em Biophysical Journal}, 66(1):259–267, 1994.

\bibitem{berger2014kolmogorov}
Vance~W Berger and YanYan Zhou.
\newblock Kolmogorov--smirnov test: Overview.
\newblock {\em Wiley statsref: Statistics reference online}, 2014.

\bibitem{descoteaux_2015}
Maxime Descoteaux.
\newblock High angular resolution diffusion imaging (hardi).
\newblock {\em Wiley Encyclopedia of Electrical and Electronics Engineering},
  page 1–25, 2015.

\bibitem{engelke2011visual}
Ulrich Engelke, Hagen Kaprykowsky, Hans-J{\"u}rgen Zepernick, and Patrick
  Ndjiki-Nya.
\newblock Visual attention in quality assessment.
\newblock {\em IEEE Signal Processing Magazine}, 28(6):50--59, 2011.

\bibitem{fadnavis_p2s}
Shreyas Fadnavis, Joshua Batson, and Eleftherios Garyfallidis.
\newblock Patch2self: Denoising diffusion {MRI} with self-supervised learning.
\newblock In H.~Larochelle, M.~Ranzato, R.~Hadsell, M.~F. Balcan, and H.~Lin,
  editors, {\em Advances in Neural Information Processing Systems}, volume~33,
  pages 16293--16303. Curran Associates, Inc., 2020.

\bibitem{gretton12a}
Arthur Gretton, Karsten~M. Borgwardt, Malte~J. Rasch, Bernhard Sch{{\"o}}lkopf,
  and Alexander Smola.
\newblock A kernel two-sample test.
\newblock {\em Journal of Machine Learning Research}, 13(25):723--773, 2012.

\bibitem{hansen2016data}
Brian Hansen and Sune~N{\o}rh{\o}j Jespersen.
\newblock Data for evaluation of fast kurtosis strategies, b-value optimization
  and exploration of diffusion mri contrast.
\newblock {\em Scientific data}, 3(1):1--5, 2016.

\bibitem{hutchinson2017analysis}
Elizabeth~B Hutchinson, Alexandru~V Avram, M~Okan Irfanoglu, C~Guan Koay,
  Alan~S Barnett, Michal~E Komlosh, Evren {\"O}zarslan, Susan~C Schwerin,
  Sharon~L Juliano, and Carlo Pierpaoli.
\newblock Analysis of the effects of noise, dwi sampling, and value of assumed
  parameters in diffusion mri models.
\newblock {\em Magnetic resonance in medicine}, 78(5):1767--1780, 2017.

\bibitem{jensen_helpern_2010}
Jens~H. Jensen and Joseph~A. Helpern.
\newblock {MRI} quantification of non-gaussian water diffusion by kurtosis
  analysis.
\newblock {\em NMR in Biomedicine}, 23(7):698–710, 2010.

\bibitem{jones2004effect}
Derek~K Jones.
\newblock The effect of gradient sampling schemes on measures derived from
  diffusion tensor mri: a monte carlo study.
\newblock {\em Magnetic Resonance in Medicine: An Official Journal of the
  International Society for Magnetic Resonance in Medicine}, 51(4):807--815,
  2004.

\bibitem{koay_ozarslan_basser_2009}
Cheng~Guan Koay, Evren Özarslan, and Peter~J. Basser.
\newblock A signal transformational framework for breaking the noise floor and
  its applications in {MRI}.
\newblock {\em Journal of Magnetic Resonance}, 197(2):108–119, 2009.

\bibitem{kubicki2007review}
Marek Kubicki, Robert McCarley, Carl-Fredrik Westin, Hae-Jeong Park, Stephan
  Maier, Ron Kikinis, Ferenc~A Jolesz, and Martha~E Shenton.
\newblock A review of diffusion tensor imaging studies in schizophrenia.
\newblock {\em Journal of psychiatric research}, 41(1-2):15--30, 2007.

\bibitem{novikov_fieremans_jespersen_kiselev_2018}
Dmitry~S. Novikov, Els Fieremans, Sune~N. Jespersen, and Valerij~G. Kiselev.
\newblock Quantifying brain microstructure with diffusion {MRI}: Theory and
  parameter estimation.
\newblock {\em NMR in Biomedicine}, 32(4), 2018.

\bibitem{novikov2018modeling}
Dmitry~S Novikov, Valerij~G Kiselev, and Sune~N Jespersen.
\newblock On modeling.
\newblock {\em Magnetic resonance in medicine}, 79(6):3172--3193, 2018.

\bibitem{ozarslan2013mean}
Evren {\"O}zarslan, Cheng~Guan Koay, Timothy~M Shepherd, Michal~E Komlosh,
  M~Okan {\.I}rfano{\u{g}}lu, Carlo Pierpaoli, and Peter~J Basser.
\newblock Mean apparent propagator (map) mri: a novel diffusion imaging method
  for mapping tissue microstructure.
\newblock {\em NeuroImage}, 78:16--32, 2013.

\bibitem{coupe_manjon_robles_collins_2012}
Coupe P., Manjon J.v., M.~Robles, and D.l. Collins.
\newblock Adaptive multiresolution non-local means filter for three-dimensional
  magnetic resonance image denoising.
\newblock {\em IET Image Processing}, 6(5):558, 2012.

\bibitem{pasternak2009free}
Ofer Pasternak, Nir Sochen, Yaniv Gur, Nathan Intrator, and Yaniv Assaf.
\newblock Free water elimination and mapping from diffusion mri.
\newblock {\em Magnetic Resonance in Medicine: An Official Journal of the
  International Society for Magnetic Resonance in Medicine}, 62(3):717--730,
  2009.

\bibitem{poldrack2016phenome}
Russell~A Poldrack, Eliza Congdon, William Triplett, KJ~Gorgolewski,
  KH~Karlsgodt, JA~Mumford, FW~Sabb, NB~Freimer, ED~London, TD~Cannon, et~al.
\newblock A phenome-wide examination of neural and cognitive function.
\newblock {\em Scientific data}, 3(1):1--12, 2016.

\bibitem{ramdas2017wasserstein}
Aaditya Ramdas, Nicol{\'a}s~Garc{\'\i}a Trillos, and Marco Cuturi.
\newblock On wasserstein two-sample testing and related families of
  nonparametric tests.
\newblock {\em Entropy}, 19(2):47, 2017.

\bibitem{rokem_2016}
Ariel Rokem.
\newblock Stanford hardi surfaces, Oct 2016.

\bibitem{Sjolund2018}
Jens Sj{\"o}lund, Anders Eklund, Evren {\"O}zarslan, Magnus Herberthson, Maria
  B{\aa}nkestad, and Hans Knutsson.
\newblock Bayesian uncertainty quantification in linear models for diffusion
  {MRI}.
\newblock {\em NeuroImage}, 175:272--285, 2018.

\bibitem{Sjolund2015}
Jens Sj{\"o}lund, Filip Szczepankiewicz, Markus Nilsson, Daniel Topgaard,
  Carl-Fredrik Westin, and Hans Knutsson.
\newblock Constrained optimization of gradient waveforms for generalized
  diffusion encoding.
\newblock {\em Journal of magnetic resonance}, 261:157--168, 2015.

\bibitem{smith2006tract}
Stephen~M Smith, Mark Jenkinson, Heidi Johansen-Berg, Daniel Rueckert, Thomas~E
  Nichols, Clare~E Mackay, Kate~E Watkins, Olga Ciccarelli, M~Zaheer Cader,
  Paul~M Matthews, et~al.
\newblock Tract-based spatial statistics: voxelwise analysis of multi-subject
  diffusion data.
\newblock {\em Neuroimage}, 31(4):1487--1505, 2006.

\bibitem{tax2022s}
Chantal~MW Tax, Matteo Bastiani, Jelle Veraart, Eleftherios Garyfallidis, and
  M~Okan Irfanoglu.
\newblock What’s new and what’s next in diffusion mri preprocessing.
\newblock {\em NeuroImage}, 249:118830, 2022.

\bibitem{tournier_calamante_connelly_2007}
J-Donald Tournier, Fernando Calamante, and Alan Connelly.
\newblock Robust determination of the fibre orientation distribution in
  diffusion {MRI}: Non-negativity constrained super-resolved spherical
  deconvolution.
\newblock {\em NeuroImage}, 35(4):1459–1472, 2007.

\bibitem{veraart_novikov_christiaens_ades-aron_sijbers_fieremans_2016}
Jelle Veraart, Dmitry~S. Novikov, Daan Christiaens, Benjamin Ades-Aron, Jan
  Sijbers, and Els Fieremans.
\newblock Denoising of diffusion {MRI} using random matrix theory.
\newblock {\em NeuroImage}, 142:394–406, 2016.

\bibitem{wang2011reduced}
Zhou Wang and Alan~C Bovik.
\newblock Reduced-and no-reference image quality assessment.
\newblock {\em IEEE Signal Processing Magazine}, 28(6):29--40, 2011.

\bibitem{wang2004image}
Zhou Wang, Alan~C Bovik, Hamid~R Sheikh, and Eero~P Simoncelli.
\newblock Image quality assessment: from error visibility to structural
  similarity.
\newblock {\em IEEE transactions on image processing}, 13(4):600--612, 2004.

\bibitem{westin2016q}
Carl-Fredrik Westin, Hans Knutsson, Ofer Pasternak, Filip Szczepankiewicz,
  Evren {\"O}zarslan, Danielle van Westen, Cecilia Mattisson, Mats Bogren,
  Lauren~J O'Donnell, Marek Kubicki, et~al.
\newblock Q-space trajectory imaging for multidimensional diffusion mri of the
  human brain.
\newblock {\em Neuroimage}, 135:345--362, 2016.

\bibitem{zhang2012noddi}
Hui Zhang, Torben Schneider, Claudia~A Wheeler-Kingshott, and Daniel~C
  Alexander.
\newblock Noddi: practical in vivo neurite orientation dispersion and density
  imaging of the human brain.
\newblock {\em Neuroimage}, 61(4):1000--1016, 2012.

\end{thebibliography}
\end{document}